\documentclass[11pt]{article}
\usepackage{amssymb}
\usepackage{amsmath}
\usepackage{amsthm}
\usepackage{amsfonts}
\usepackage{url}
\usepackage{hyperref}
\usepackage{breakurl}
\begin{document}
\bibliographystyle{plain}
\title{{Some Comments on C. S. Wallace's Random Number Generators}%
\thanks{This paper was originally dedicated to 
Professor Chris Wallace~\cite{wiki} on the 
occasion of his 70th birthday, but unfortunately Chris passed away before
it could be published. Another version appeared in
{\em The Computer Journal}, {\bf{51}}, 5 (2008), 579--584.}
\thanks{Copyright \copyright 2003--2010 R.~P.~Brent \hspace*{\fill} rpb213}
}
\author{~\\[10pt]Richard P.\ Brent\\
        Mathematical Sciences Institute\\
        Australian National University\\
        Canberra, ACT 0200, Australia\\[30pt]
In Memory of Christopher Stewart Wallace 1933--2004} %
\date{}

\newcommand{\E}{{\rm{E}}}
\newcommand{\mybar}{\overline}
\newcommand{\rpb}{{R. P. Brent}}
\newcommand{\realtilde}{\makebox{\tt\char'176}}
\newcommand{\lbrk}{{\linebreak[0]}}
\newcommand{\MC}{{Math.\ Comp.\ }}
\newcommand{\LNCS}{{Lecture Notes in Computer Science}}
\newcommand{\mmod}{{\;\bmod\;}}
\newcommand{\dhat}{{\widehat{d}}}

\maketitle

\begin{abstract}
We outline some of Chris Wallace's contributions to pseudo-random
number generation.  In particular, we consider his recent idea for
generating normally distributed variates without relying on a source
of uniform random numbers, and compare it with more conventional methods
for generating normal random numbers.  Implementations of Wallace's idea
can be very fast (approximately as fast as good uniform generators).
We discuss the statistical quality of the output, and mention 
how certain pitfalls can be avoided.
\end{abstract}

\thispagestyle{empty}

\section{Introduction}
\label{sec:intro}

In many simulation, graphics, simulated-annealing, cryptographic 
and Monte Carlo/Las Vegas programs, 
a substantial fraction of the time is used in generating
pseudo-random numbers from the uniform, normal or other distributions,
so methods of generating such numbers have received much \hbox{attention}.

This paper is dedicated to the memory of Chris Wallace, 
and our intention is to outline
Wallace's substantial contribution to several aspects of random number
generation, both in hardware and software.  

In \S\ref{sec:hardware} we consider hardware random number generators (RNGs),
and in \S\ref{sec:uniform} we mention (software) uniform RNGs. 
In \S\ref{sec:normal} we consider ``conventional'' normal
random number generators, and in \S\ref{sec:fastnorm} we consider
Wallace's new ``maximum entropy'' idea for normal RNGs
that do not depend in an essential way on a source of uniform RNGs.
This idea is aesthetically appealing (why bother to
generate uniform random numbers just in order to transform them by some
time-consuming process into normal random numbers?) and has the potential
to give extremely fast normal RNGs.  

\section{Hardware RNGs}
\label{sec:hardware}

In some cryptographic applications it is important for the numbers to be
genuinely random, in the sense of being unpredictable,
and not merely ``pseudo-random'', in the sense of passing various
statistical tests. For example, this is
the case when generating ``one-time pads'', 
or when constructing random primes
whose products are to be made public for use with the Rivest, Shamir and
Adleman ``RSA'' public-key cryptosystem~\cite{RSA},
or when constructing exponents to be used in the Diffie-Hellman key
exchange protocol or the El~Gamal public-key cryptosystem~\cite{MOV}.

Wallace in~\cite{Wallace90}
described a simple hardware device that could provide a stream of
unpredictable 32-bit numbers at a rate of 64 Mbit/sec, using a 4~MHz clock.
The device was connected to the memory-mapped I/O bus of a multiprocessor
computer, and appeared to a software process as a single 32-bit word of memory
whose content was different (and unpredictable) every time it was read.
Technology has advanced since 1990 so an implementation using similar ideas
could now use a much faster clock.
This could, for example, be used in the implementation of 
Rabin's ``everlasting encryption'' scheme~\cite{ADR,AR,Maurer92,Rabin83}, 
which depends on the availability of a high-volume stream of random 
and unpredictable bits.

\section{Uniform RNGs}
\label{sec:uniform}

In~\cite{Wallace89}, Wallace considered several ways of obtaining
uniform pseudo-random number generators with period close to $2^{64}$
on machines with 32-bit words.  There are many ways to accomplish 
this~\cite{rpb132,Ferrenberg92,Knuth,Marsaglia85,Marsaglia91}, 
but most of them require a large amount of state 
information. This can be a problem if several independent streams
of random numbers are required simultaneously. With Wallace's proposal,
only two 32-bit words of state information are required.

\section{Conventional Normal RNGs}
\label{sec:normal}

Most popular algorithms for generating normally distributed pseudo-random
numbers are based on some variant of the rejection method, pioneered by 
von Neumann~\cite{vonNeumann}.
More recent references 
are~\cite{AD72,rpb023,Devroye,Forsythe,Kinderman,Leva92}.
Wallace~\cite{Wallace76} contributed some elegant and 
efficient generators of this class.

Rejection methods for normally distributed pseudo-random numbers require on
average some number $U > 1$ of uniformly distributed numbers per normally
distributed number. Thus, they can not be faster than the uniform random
number generator, and are typically several times slower. 
Rejection methods for the normal distribution usually (though not
always~\cite{rpb023,Forsythe}) involve the computation of functions
such as $\log$, $\sin$, $\cos$, which is slow compared to the time 
required to generate a uniform pseudo-random number.
Leva~\cite[Table~1]{Leva92} compared several of the better methods
and found that they are at least five times slower than a fast uniform 
generator on the same machine.

\section{Maximum Entropy Normal RNGs}
\label{sec:fastnorm}

Wallace~\cite{Wallace96} revolutionised 
normal random number generation by his discovery of a 
class of methods that do not depend in an essential way
on uniform generators. 
Similar ideas can be used to generate pseudo-random numbers
with some other distributions.
In Wallace's paper~\cite{Wallace96} the uniform, Gaussian (normal)
and exponential distributions are considered as maximum-entropy 
distributions subject to the following constraints:

\begin{enumerate}
\item[] Uniform: $0 \le x \le 1$
\item[] Gaussian: $\E(x^2) = 1$
\item[] Exponential: $\E(x) = 1$, $x \ge 0$.
\end{enumerate}

The idea of a maximum-entropy distribution is most easily seen
in the discrete case of $N$ possibilities with 
probabilities $p_1, \ldots, p_N$.  Subject to the
constraints $p_j \ge 0$ and $\sum p_j = 1$, the 
uniform distribution $p_1 = \cdots = p_N = 1/N$
maximises the entropy $S = -\sum p_j \log p_j$.
This can be proved using Lagrange multipliers.
Similarly,
the continuous distribution on $[0,1]$ that maximises
$-\int_0^1 f(x) \log f(x) dx$ is the uniform distribution,
and the continuous distribution on $(-\infty,+\infty)$ that
maximises $-\int_{-\infty}^{+\infty} f(x) \log f(x) dx$
subject to $\int_{-\infty}^{+\infty} x^2 f(x) dx = 1$ is the
Gaussian distribution. These statements can be proved using the
calculus of variations.
For the reader unfamiliar with
Bayesian and maximum entropy methods, 
a good introduction is Jaynes~\cite{Jaynes1986a}.
An annotated bibliography is available at~\cite{Jaynes1994}.

In the following we restrict our attention to the Gaussian case, since
that is where Wallace's idea gives the most significant speedup over
conventional methods. For example, 
Wallace's own implementation {\em FastNorm} is reported
in~\cite[\S5]{Wallace96} to be only 13~percent slower than a 
generalised Fibonacci uniform random number generator on a RISC workstation.

Wallace proposed his method in a Technical Report in 1994, and a revision
of this Report appeared two years later~\cite{Wallace96} along with
an implementation {\em fastnorm}.
Some changes
in the implementation were made in 1998, %
resulting in an improved implementation {\em fastnorm2}~\cite{fastnorm2}.
There is a more recent and probably better implementation 
{\em fastnorm3}~\cite{fastnorm3}, 
but it was not available when our tests were performed,
so we restrict our comments to {\em fastnorm2}.

Wallace~\cite{Wallace96} describes two implementations~-- one using
integer arithmetic, and the other using floating-point arithmetic.
On the workstation that he tested it on, the integer version was
faster, but this might not be true on more recent machines with
faster floating-point hardware.

Traditional normal RNGs are inefficient on vector processors.
In 1993 the author compared various normal RNGs on vector 
processors and concluded that careful implementations of old 
methods such as the 1958 {\em Polar} method of
Box, Muller and Marsaglia (see Knuth~\cite[Algorithm~P]{Knuth})
and the 1959 Box-Muller method~\cite{Knuth,Muller} were faster than
more recent methods~\cite{Leva92} on vector processors produced by
companies such as Cray, Fujitsu, and NEC:
see~\cite{rpb141,Petersen88}.
When Wallace's maximum-entropy idea appeared, it was clear that the
landscape had changed, although the published implementation 
{\em fastnorm} was not intended to be efficient on vector processors.
Thus, the author implemented an efficient vectorised
version {\em rann4}~\cite{rpb170,rpb185} 
of Wallace's maximum-entropy idea.
{\em rann4} and a more recent implementation {\em rannw}~\cite{rannw} are 
more than three times faster than the methods previously
thought to be the most efficient on vector processors.

\subsection{Wallace's {\em fastnorm} algorithm}

Many uniform random number generators generate one or more new uniform
variables from a set of previously-generated uniform variables. 
Wallace's idea is to
apply the same principle to normal random number generators. Given a set of
normally distributed random variables, we can generate a new set of
normally distributed random variables by applying a linear
transformation that respects the ``maximum entropy'' constraint.
This avoids the time-consuming conversion of uniform to normal variables
that is required in conventional normal random number generators
(see \S\ref{sec:normal}).

The key idea is: if $x$ is an $n$-vector of
independent, identically distributed $N(0,1)$ random variables
$x_1, \ldots, x_n$, 
and $Q$ is any $n \times n$ orthogonal matrix,
then $y = Qx$ is another $n$-vector of independent, 
identically distributed $N(0,1)$ random variables.
(Of course, the components $y_i$ of $y$ are dependent on the components
$x_j$ of $x$.)
To prove the claim, observe that the component $x_j$ has probability density
$(2/\pi)^{-1/2}\exp(-x_j^2/2)$, so the vector $x$ has probability density
$(2/\pi)^{-n/2}\exp(-r^2/2)$, where $r = \Vert x \Vert_2$.
This density depends only on $r$, the distance of $x$ from the origin.
However, since $Q$ is orthogonal, 
$\Vert y \Vert_2 =  \Vert Qx \Vert_2 = \Vert x \Vert_2 = r$.

Suppose that the $n$-vector $x$ is a pool of $n$ pseudo-random numbers 
that (we hope) are independent and normally distributed.  
We can generate a new pool $y = Qx$ by
applying an orthogonal transformation~$Q$.  
However, several problems arise.

\subsection{Undesirable correlations}
\label{subsec:correlations}
$y_i$ is correlated with $x_j$.  In fact,
$y_i = q_{i,j}x_j + \cdots$, so
$\E(y_ix_j) = \E(q_{i,j}x_j^2) = q_{i,j}$.
This problem can be overcome by applying several different orthogonal
tranformations $Q_1, Q_2, \ldots$
with a random choice of signs, so when averaged over all transformations 
$\E(q_{i,j}) \approx 0$.

\subsection{Cost of transformations} 
It is too expensive to apply a general $n \times n$ orthogonal
transformation $Q$ to produce $n$ new random numbers.  This would involve
of order $n$ multiplications (and a similar number of additions)
per random number generated.  To overcome this problem, we can take
$Q$ to have a special form, e.g.~in {\em rann4} we use a product of 
plane rotations of the form 
\[
R(\theta) = \left[\begin{array}{cc}
		\cos \theta	& \sin \theta \\
		-\sin \theta	& \cos \theta
		\end{array} \right]\;,
\]
where $\theta$ varies, but is held constant within each inner loop.
We do not need to compute trigonometric functions, since
$\sin \theta = 2t/(1+t^2)$ and
$\cos \theta = (1-t^2)/(1+t^2)$, where
$t = \tan (\theta/2)$ varies; the angle $\theta$ is defined only for 
mathematical convenience and is never computed.

In his implementation {\em fastnorm}, 
Wallace preferred to use $4 \times 4$ orthogonal
matrices $A_1, A_2, A_3, A_4$, where
\[
A_1 = \frac{1}{2}\left[\begin{array}{rrrr}
	1 	& 1	& -1	& 1 \\
	1	& -1	& 1	& 1 \\
	1	& -1	& -1	& -1 \\
	-1	& -1	& -1	& 1
	\end{array} \right]\;,  
\]
and $A_2, A_3, A_4$ are similar.
The advantage (on a machine with slow floating-point multiplication) is
that multiplication of a $4$-vector by $A_1$ requires only seven
additions and one division by two (for details see~\cite[\S2.1]{Wallace96}).

The inner loop of the implementation
is similar to the inner loop for the popular ``generalised Fibonacci''
uniform random number 
generators~\cite{Anderson,rpb132,Green,%
James,Knuth,Marsaglia85,Petersen94,Reiser}.
Wallace's implementation of {\em fastnorm} on a RISC workstation is
about as fast as a good uniform random number generator on the same
workstation.

\subsection{Mixing}
\label{subsec:mixing}
As Wallace observes~\cite[\S2.2]{Wallace96}, it is desirable that
any value in the pool should eventually contribute to every value in the
pools formed after several passes. In other words, the transformation
from one pool to the next should be strongly ``mixing''.  
In our experience this
is a tricky aspect of the implementation of generators based on Wallace's
idea~-- several attempts which appeared plausible did not produce acceptable
random numbers (after transformation to uniform variates they failed
various statistical tests in Marsaglia's Diehard package~\cite{Diehard}).

In {\em fastnorm}, Wallace ensures mixing by regarding the pool of 1024
values as a $256 \times 4$ matrix which is (implicitly) transposed at
each pass; an additional ad hoc permutation is applied by stepping some row
indices with an odd stride (mod~256). %
For details see~\cite[\S2.2]{Wallace96}.

In {\em rann4} we effectively apply permutations of the form
$\pi_1(j) = \alpha j + \gamma \mmod n$,
$\pi_2(j) = \beta  j + \delta \mmod n$,
where $\gcd(\alpha,n) = \gcd(\beta,n) = 1$. Since $n$ is a power of~2,
any odd $\alpha$ and $\beta$ can be chosen. For details 
see~\cite[\S3]{rpb170}.

Although the mixing transformations used in {\em fastnorm} and {\em rann4}
appear satisfactory, they seem {\em ad hoc} and there is little helpful theory  
here~-- all we can do is apply empirical tests.

\subsection{Chi-squared correction}
Because $Q$ is orthogonal, $\Vert Qx \Vert_2 = \Vert x \Vert_2$,
so the sum of squares of numbers in a pool remains constant. This is
unsatisfactory, because if $x_1, \ldots, x_n$ were independent samples from
the normal $N(0,1)$ distribution, we would expect
$\sum_{1 \le i \le n}x_i^2$ to have a chi-squared distribution
$\chi_\nu^2$, where $\nu = n$ is the pool size.

To overcome this defect,
Wallace suggests that one pseudo-random number from each pool
should not be returned to the user, but should be used to
approximate a random sample $S$ from the $\chi^2_\nu$ distribution.
A scaling factor can be introduced to ensure
that the sum of squares of the $\nu$ values in the pool (of which $\nu-1$
are returned to the user) is~$S$.  
If the routine is written to provide random numbers with
mean $\mu$ and variance $\sigma^2$, %
then scaling by $S^{1/2}$ can be done at the same time as
scaling by $\sigma$, so it is essentially free.

There are several approximations to the $\chi^2_\nu$ distribution
for large $\nu$. For example, the one used in {\em rann4} is
\[
2\chi^2_\nu \;\simeq\; \left(x + \sqrt{2\nu - 1}\right)^2\;,
\]
where $x$ is $N(0,1)$.
It would not be much more expensive to use the (more accurate)
Wilson-Hilferty approximation~\cite{Wilson}
\[
\chi^2_\nu \;\simeq\; \nu\;\left(\left(\frac{2}{9\nu}\right)^{1/2}x \;+\; 
		\left(1 - \frac{2}{9\nu}\right)\right)^3\;.
\]
Even better is 				%
\[
\chi^2_\nu \;\simeq\; A(x^2 - 1) \;+\; (2(\nu - A^2))^{1/2}\;x \;+\; \nu\;,
\]            
where
\[
A = 2\sqrt{\nu}\;\sin\left(\frac{1}{3}\arcsin\frac{1}{\sqrt{\nu}}\right) 
= \frac{2}{3} + 
  O\left(\frac{1}{\nu}\right)    %
\]
satisfies the cubic equation $A^3 - 3\nu A + 2\nu = 0$.
We can assume that $\nu$ is large ($\nu = 1024$ in {\em fastnorm};
$\nu$ depends on the size of the buffer provided by the user
in {\em rann4/rannw}),
so all of these approximations are sufficiently accurate. 
A slow but exact $\chi_\nu^2$ algorithm, such as that of
Ahrens and Dieter~\cite{AD82}, is not required.

In the above approximations to $\chi_\nu^2$, the variable $x$ was 
supposed to have a normal distribution. If only $n-1$ values are
returned to the user from a pool of $n$ values, the remaining (scaled)
value $x$ can be used to approximate $\chi_\nu^2$ for the next pool.
This is a point where the implementations of {\em fastnorm} and
{\em fastnorm2} differ. In {\em fastnorm}, $x$ is taken from the 
current pool, but in {\em fastnorm2} it is taken from the previous
pool. The choice used in {\em fastnorm} is undesirable because a
large value of $x$, and hence a large scaling factor from the
$\chi_\nu^2$ approximation, is correlated with a small sum of
squares of the remaining values in the pool (since the sum of
squares including $x$ is invariant).

\subsection{More subtle correlations}
\label{subsec:squared}

In \S\ref{subsec:correlations} we saw how, by using several orthogonal
transformations, we could ensure that
$\E(y_ix_j) \approx 0$. However, more subtle correlations persist.
Consider the simplified model
\[
\left[\begin{array}{c} y_1 \\ y_2 \end{array}\right]
	= R(\theta) \left[\begin{array}{c} x_1 \\ x_2 \end{array}\right]\;,
\]
where $R(\theta)$ is a plane rotation as above, and $\theta$ is
distributed uniformly in $[0, 2\pi)$.  We write $c = \cos \theta$,
$s = \sin \theta$.  Thus
$y_1 = cx_1 + sx_2$, $y_2 = -sx_1 + cx_2$.
Suppose that $x_1$ and $x_2$ are independent and normally distributed,
with zero mean and unit variance.
Then
\[
E(x_1^2 y_1^2) = E(c^2)E(x_1^4) + E(s^2)E(x_1^2 x_2^2)
	= 2 \ne E(x_1^2)E(y_1^2)\;.
\]
In {\em fastnorm/fastnorm2} and {\em rann4/rannw}, 
similar effects occur,
although the undesirable correlations are small and they occur between
well-separated outputs (the separations are of the order of the pool size) 
because of the
permutations used to provide mixing ({\em{cf}}~\S\ref{subsec:mixing}).

\subsection{Other finite pool size effects}
\label{subsec:finite}

Chris Wallace~\cite{trouble} has observed a phenomenon that, 
like the one discussed in~\S\ref{subsec:squared}, becomes less
significant as the pool size increases, but never disappears entirely
for any finite pool size~$n$.

Consider a rare event such as the occurrence of a large normal variate
$x$ that is expected to occur say once in every $10n$ samples,
i.e.~once in every 10 pools. The ``energy'' $x^2$ is distributed over
only a small number (four) of variables in the next pool.
Thus we can expect one or more of these variables to be unusually
large.  Although the distribution of values considered over many pools
is correct, it is more likely that rare events will occur in adjacent pools.

It is possible %
to devise statistical tests that \hbox{detect} this behaviour
and/or the correlations described in~\S\ref{subsec:squared}.  However, 
we have not obtained any statistically significant \hbox{results} 
with a sample size of less than ${10^4}n$. 

Clearly, one way to reduce (though not eliminate) the significance
of such effects is to increase the pool size (easy for {\em rann4/rannw}).
Another way is to discard some of the numbers produced by the random
number generator~-- e.g.~we could use every third value, or the values
in every third pool.  This has an obvious effect on the speed of the
generator, but because the underlying algorithm is so fast we can
afford to do it and still have a random number generator that is
faster than more conventional generators ({\em{cf}}~\S\ref{sec:normal}).

\subsection{Use of uniform RNGs}
Although normal generators based on the maximum entropy idea do not
use uniform random numbers in any {\em essential} way, it is convenient
to use a uniform RNG for purposes such as initialisation, selection
of orthogonal transformations, etc.
The advantage of the maximum entropy methods
is that the number $U$ of uniform distributed numbers 
required per normally distributed number is very small (of the order
of $1/n$ for pool size $n$), whereas for rejection methods $U > 1$.

If we choose a uniform random number generator with known long period,
and use it at least once for each pool of normal random numbers
(e.g. to select from a set of possible orthogonal transformations),
then it is easy to guarantee that the period of the normal random
number generator is at least as great as that of the uniform random
number generator.  Thus, although any use of a uniform random number
generator might be considered contrary to the spirit of the maximum
entropy method, it does have the practical benefit of guaranteeing a 
long period.  If (as is certainly possible) we avoided using a uniform
generator except perhaps for initialisation, then we could not 
{\em guarantee} a long period, although a short period would 
be extremely unlikely, since it would require an implausible
coincidence in the initialisation.

\subsection{Summary}

Although care needs to be taken in the implementation of normal random number
generators like {\em fastnorm}, and the end-user should be aware of
the small but unavoidable \hbox{defects} discussed in
\S\S\ref{subsec:squared}-\ref{subsec:finite}, these generators have
such a performance advantage over more conventional generators
that they can not be ignored in applications where the speed of
generation of pseudo-random numbers is critical. 

\section*{Acknowledgements}

Comments by David Dowe and by Chris Wallace himself,
on earlier versions of this paper, are gratefully acknowledged.

\end{document}